\documentclass[twocolumn,showpacs,preprintnumbers,amsmath,amssymb]{revtex4}
\usepackage{graphicx,amsmath,amssymb,bm,makeidx}
\begin{document}

\title{Magnetic quantum oscillations in doped antiferromagnetic insulators}
\author{V. V. Kabanov$^{1,2}$ and A. S. Alexandrov$^2$ }

\affiliation{$^1$Jozef Stefan Institute 1001, Ljubljana, Slovenia\\
$^{2}$Department of Physics, Loughborough University, Loughborough
LE11 3TU, United Kingdom}

\begin{abstract}
Energy spectrum of electrons (holes) doped into a two-dimensional
antiferromagnetic insulator is quantized  in an external magnetic
field of arbitrary direction. A peculiar dependence of de Haas-van
Alphen (dHvA) or Shubnikov-de Haas (SdH) magneto-oscillation
amplitudes on
 the azimuthal in-plane angle from the magnetization direction
and on the polar angle from the out-of-plane direction is found,
which can be used as a sensitive probe of the antiferromagnetic
order in doped Mott-Hubbard, spin-density wave (SDW), and
conventional band-structure  insulators.

\end{abstract}
 \maketitle

Quantum oscillations of  magnetization and resistivity with the
magnetic field are of a great experimental and theoretical value
providing reliable and detailed Fermi-surfaces \cite{shoen,sin,kar}.
Specifically interest in dHvA and SdH effects in almost
two-dimensional (2D) Fermi-liquids has recently gone through a
vigorous revival due to experimental discoveries of
magneto-oscillations  in a few high-temperature cuprate
superconductors \cite{ley,yel,ban,proust}. Their description in the
framework of the standard theory for a metal \cite{shoen} has led to
a  small electron-like Fermi-surface area of a few percent of the
first Brillouin zone and to a surprisingly
 low Fermi energy of  about the room temperature
 \cite{proust,yel},  somewhat inconsistent with the first-principle
(LDA) band structures and angle-resolved photoemission (ARPES)
spectra of cuprates \cite{shen}. The oscillations have been observed
in the \emph{superconducting} (vortex) state well below the upper
critical field raising a doubt concerning their normal state origin
\cite{aleosc}. While a better understanding of dHvA/SdH effects in
doped antiferromagnetic insulators is generally important, it
becomes particularly vital for building an adequate theory of
high-temperature superconductivity since parent cuprates are
antiferromagnets.

Here, using a tight-binding Hamiltonian  we  quantize the energy
spectrum of electrons or holes moving on the anti-ferromagnetic (AF)
background in a two dimensional lattice. We find a peculiar
dependence of the magneto-oscillation amplitudes on the magnetic
field direction, which could serve as a sensitive probe of the
antiferromagnetic order in doped insulators.

The mean-field tight-binding Hamiltonian of carriers doped into
 the bipartite antiferromagnetic
 square lattice in the external magnetic field, ${\bf B}$, is written  as
\begin{eqnarray}
H&=&\sum_{ii'} \delta_{ii'} (\Delta
\hat{a}^{\dagger}_{i}\sigma_{z}\hat{a}_{i}+\mu_{B}\bm {B}
\hat{a}^{\dagger}_{i}\bm {\sigma}\hat{a}_{i})+ t_{ii'}
\hat{a}^{\dagger}_{i'}\hat{a}_{i} \cr &-& \sum_{jj'}
 \delta_{jj'} (\Delta
\hat{b}^{\dagger}_{j}\sigma_{z}\hat{b}_{j}- \mu_{B}\bm {B}
\hat{b}^{\dagger}_{j}\bm{\sigma} \hat{b}_{j})+ t_{jj'}
\hat{b}^{\dagger}_{j'}\hat{b}_{j} \cr &+& \sum_{ij}
t_{ij}\hat{b}^\dagger_{j}\hat{a}_{i}+H.c.,\label{ham}
\end{eqnarray}
where $\hat{a}^{\dagger}_{i}=(a^{\dagger}_{i\uparrow},
a^{\dagger}_{i\downarrow})$ and
$\hat{b}^{\dagger}_{i}=(b^{\dagger}_{i\uparrow},
b^{\dagger}_{i\downarrow})$ create the carrier on sites "$i$" and
"$j$" of sublattices $A$ and $B$, respectively, with the spin
$s=\uparrow,\downarrow$, $\Delta$ is the carrier spin-lattice spin
exchange energy (the antiferromagnetic gap), $\delta_{ii'}$ is the
Kroneker symbol, $t_{ii'}, t_{jj'}$ and $t_{ij}$ are the hopping
integrals, and $\bm {\sigma}\equiv \{ \sigma_x, \sigma_y, \sigma_z
\}$ are the Pauli matrices.

Fourier transforming the operators from Wannier (site) to Bloch
(momentum, {\bf k}) representation and assuming translational
invariance the carrier energy spectrum, $E(\bf k)$, is found by
diagonalizing $4\times4$ matrix:
\begin{eqnarray}
\left (\begin{array}{cc}t^\prime_{\bf k} -\Delta\sigma_{0} -
\mu_{B}B_{\parallel}\sigma_{z} & \mu_{B}B_{\perp}\sigma_{0}+
t_{\mathbf{k}}\sigma_{x}\\ \mu_{B}B_{\perp}\sigma_{0}+
t_{\mathbf{k}}\sigma_{x} & t^\prime_{\bf k}+\Delta\sigma_{0} +
\mu_{B}B_{\parallel}\sigma_{z} \end{array} \right ),
\end{eqnarray}
where $t^\prime_{\bf k}=\sum_{jj'} t_{jj'} \exp(i[\bf k \cdot
(j'-j)])$ is the hopping energy within one sublattice, $t_{\bf
k}=\sum_{ij} t_{ij} \exp(i[\bf k \cdot (i-j)])$ is the
inter-sublattice hopping energy. This matrix corresponds to the
choice of the 4-dimensional vector in the spin and sublattice space
at fixed $\mathbf{k}$.
\begin{figure}
\begin{center}
\includegraphics[angle=-90,width=0.50\textwidth]{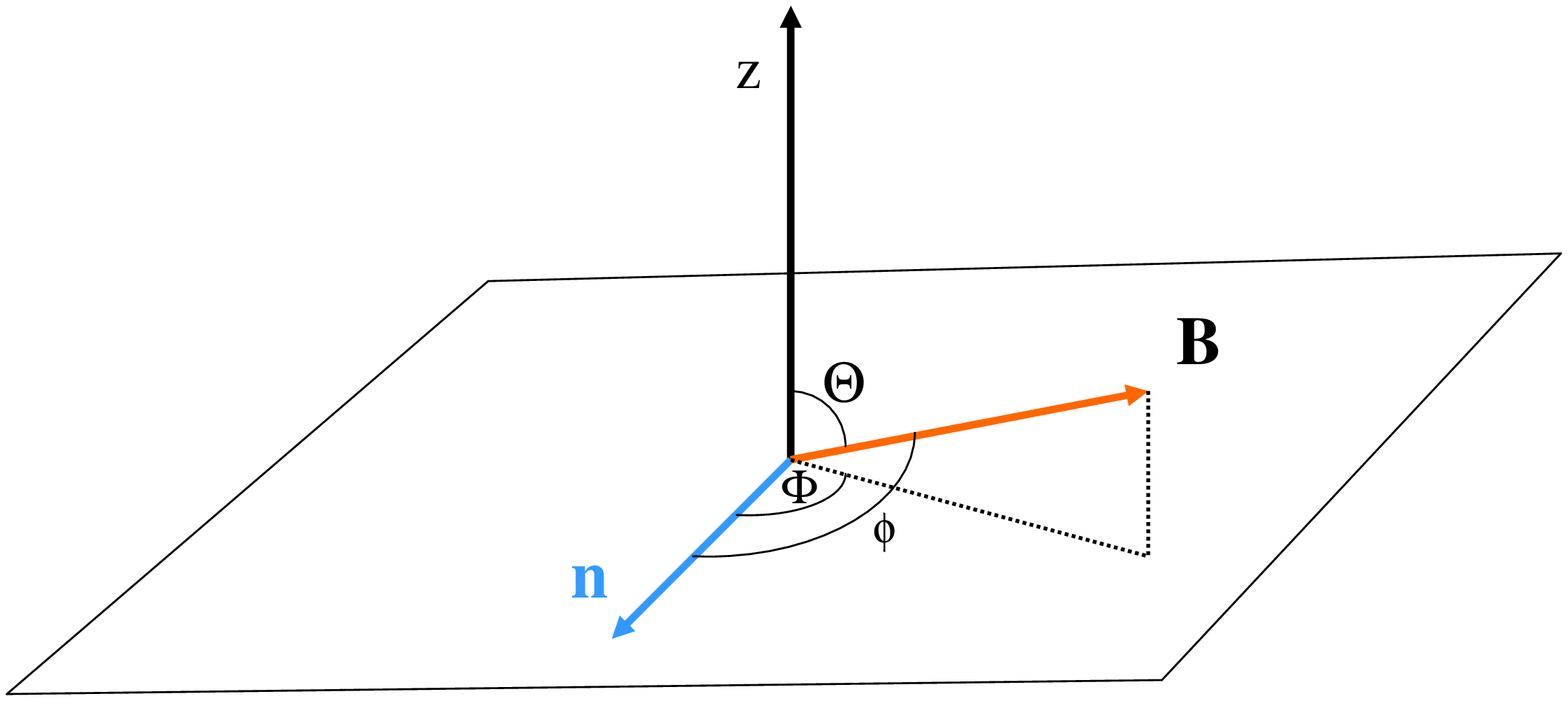}\\[0.2cm]
\caption{The azimuthal in-plane angle, $\Phi$, from the
magnetization direction, ${\bf n}$, and  the polar angle, $\Theta$,
of the magnetic field ${\bf B}$  from the out-of-plane direction.}
\end{center}
\end{figure}
Here $B_{\perp}$ and $B_{\parallel}$ are transverse and longitudinal
components of the magnetic field with respect to the lattice
magnetization ${\bf n}$, Fig.1,   $\sigma_{0}$ is the identity
matrix, and $\mu_{B}$ is the Bohr magneton. There are two electron
and two hole bands dispersed as
\begin{equation}
E({\bf k})=t^\prime_{\bf k}\pm [\Delta^2+t^2_{\bf k}]^{1/2}\pm
\mu_{B}\left [B_{\parallel}^2 +{t^2_{\bf k}\over{\Delta^2+t^2_{\bf
k}}} B_{\perp}^2 \right ]^{1/2} \label{general}
\end{equation}
in the lowest order with respect to the field, $\mu_B B\ll
\Delta$. They are split by the external magnetic field into two
subbands each with anysotropic $g$-factor, $g=2 [\cos^2(\phi) +
\sin^2(\phi) t^2_{\bf k}/(\Delta^2+t^2_{\bf k})]$ depending on the
angle $\phi$ between the field and the magnetization, Fig.1.

The anisotropic $g$-factor differs significantly from the
free-electron $g_e=2$ near the extremum points of the
valence/conductance bands, where $t^2_{\bf k}\ll \Delta^2$.
According to general principles of quantum mechanics deviations of
the $g$-factor from its classical value are related to spin-orbit
interaction. The spin-orbit interaction is not included explicitly
to the Hamiltonian Eqs.(1,2). Basically the difference originates
from the spin-orbit interaction pinning the lattice magnetization
along a crystal lattice direction and present in the Hamiltonian
implicitly. At a relatively low doping with the Fermi energy,
$E_F$ near the top (bottom) of the valence (conduction) band, one
can expand Eq.(\ref{general}) in powers of $t/\Delta$,
\begin{equation}
E({\bf k})\approx
{\hbar^2k_{x}^{2}\over{2m_{x}}}+{\hbar^2k_{y}^{2}\over{2m_{y}}}
\pm \mu_B \left[B_{\parallel}^{2}+ \gamma^{2}({\bf
k})B_{\perp}^{2}\right ]^{1/2}. \label{appr}
\end{equation}
Here $m_{x}^{-1}=4a^{2}(2t^{2}/\Delta-t^{'})/\hbar^2$,
$m_{y}^{-1}=4a^2t^{'}/\hbar^2$ are components of the effective
mass tensor, $a$ is the lattice constant, $t$ and $t^{'}>0$ are
nearest and nearest next neighbor hopping integrals, respectively,
and the
 coefficient $\gamma ({\bf k})$ is  small as $\gamma ({\bf
 k})= 2\sqrt{2}tk_{x}/\Delta \sim (E_F/\Delta)^{1/2} \ll 1$.
 Here $k_x,k_y$ are
deviations of the wave vector perpendicular and parallel to the
antiferromagnetic Brillouin zone boundary, respectively, and the
energy of the extremum point is taken as zero.

The anisotropic $g$-factor in doped antiferromagnetic insulators
was originally derived in a weak-coupling nesting model
\cite{brazov2}. Actually the effective mass approximation,
Eq.(\ref{appr}), can be also derived phenomenologically using the
symmetry arguments \cite{brazov}.  The non-unitary group of the
antiferromagnetic lattice  is $G=\{D_{4},R{\bf T}\}$, here $D_{4}$
describes all rotations which remain the system invariant.
Translations ${\bf T}$ by a lattice period  transform from one
sublattice to another changing the sign of the magnetization.
Hence, these translations are multiplied by the time inversion
operator $R$. Following Brazovskii and  Lukyanchuk  \cite{brazov}
one can construct the Hamiltonian of the required symmetry as
\begin{eqnarray}
&&H=\left({{\hbar^2k_{x}^{2}\over{2m_{x}}}}+{\hbar^2k_{y}^{2}
\over{2m_{y}}}\right) \hat{a}^{\dagger}\sigma_{0}\hat{a} + \cr &&
\mu_{B}\left[(\mathbf{B}
\cdot\mathbf{n})\hat{a}^{\dagger}(\mathbf{n}\cdot \bm
{\sigma})\hat{a}+ \gamma ({\bf k})(\mathbf{B} \times \mathbf{n})
\hat{a}^{\dagger}(\mathbf{n}\times \bm {\sigma})\hat{a} \right]
\end{eqnarray}
with  the electron (hole) energy spectrum Eq.(\ref{appr}). Here
$\mathbf{n}$ is the magnetization unit-vector, and
$\hat{a}^{\dagger}=(a^{\dagger}_{\uparrow}a^{\dagger}_{\downarrow}),\hat{a}$
are  creation and annihilation operators, respectively, for the
spinor describing the hole (electron) band. The coefficient
$\gamma({\bf k})$ is an odd function of $k_{x}$, which is zero at
the antiferromagnetic Brillouin zone boundary, so that
$\gamma({\bf k})=\gamma k_x$, where $\gamma$ does not depend on
${\bf k}$. The coupling to the magnetic field in this Hamiltonian
is obtained noticing that the transformation
$\mathbf{k}\to\mathbf{k+Q}$ with $\mathbf{Q}=\pi {\bf a}/a^{2}$ is
equivalent to the rotation in the spinor space described by the
matrix $\mathbf{n}\cdot\bm {\sigma}$ \cite{brazov} (here ${\bf
a}=\{a,a\}$). Direct comparison of the eigenvalues of the
Hamiltonian Eq.(5) with the spectrum Eq.(3) yields
$\gamma=2\sqrt{2} t/\Delta$. Importantly, the symmetry arguments
are applied beyond the mean-field approximation, Eq.(\ref{ham}),
so that spin fluctuations just renormalize the effective mass
tensor and other coefficients in Eq.(\ref{appr}).

The orbital quantization of the spectrum, Eq.(\ref{general}), is
readily obtained via the Peierls substitution \cite{peirels},  ${\bf
k} \Rightarrow -i {\bf \nabla}+e {\bf A}$  with the vector potential
${\bf A}({\bf r})$ in Eq.(2). In the lowest order with respect to
$E_F/\Delta$ we can use the effective mass approximation,
Eq.(\ref{appr}), which yields the conventional Fock-Landau levels
\cite{fock,lan}  split by the longitudinal field as
\begin{equation}
E_n = \hbar \omega |\cos(\Theta)|(n+1/2)  \pm \mu_B |B_{\parallel}|,
\label{quantum}
\end{equation}
where $\omega= eB/(m_xm_y)^{1/2}$ is the cyclotron frequency,
$n=0,1,2...$, and $\Theta$ is the polar angle between the magnetic
field and  the out-of-plane direction, Fig.1.

Now  the oscillating part of the magnetization, $\tilde{M}$, is
calculated following the standard route by applying the Poisson summation
\cite{shoen}:
\begin{equation}
\tilde{M}=\sum_{r=1}^{\infty} M_r \sin {2\pi r F\over{B}}.
\end{equation}
Here
\begin{equation}
M_r=A_r(\Theta)\cos \left[{\pi r (m_xm_y)^{1/2} \tan(\Theta) \cos
(\Phi)\over {m_e }}\right] \label{M}
\end{equation}
is the amplitude of $r$-harmonic with
\begin{eqnarray}
A_r(\Theta)&=&(-1)^{r+1}{eE_F\cos(\Theta)\over{2\pi^2 \hbar dr}} \cr
&\times& R_T\left({2\pi^2 r k_BT\over {\hbar\omega\cos(\Theta)}}
\right) R_D \left ({2\pi r \Gamma \over{\hbar\omega
\cos(\Theta)}}\right), \label{A}
\end{eqnarray}
$F=(m_xm_y)^{1/2} E_F /e\hbar \cos(\Theta)$ is the fundamental
frequency of oscillations, $R_T(z)=z/\sinh(z)$ and
$R_D(z)=\exp(-z)$ are conventional temperature and Dingle
reduction factors, $\Gamma$ is the scattering rate, $m_e$ is the
free electron mass, $d$ is the inter-plane distance, and $\Phi$ is
the azimuthal in-plane angle from the magnetization direction,
Fig.1. Both angles $\Theta$ and $\Phi$ in Eq.(\ref{M}) are
changing in the interval $0\leqslant \Theta,\Phi\leqslant \pi/2$.
Three-dimensional corrections to the energy spectrum can be
accounted for by the additional  Yamaji factor \cite{yam},
$R_Y=J_0 \left[4\pi r \tilde{t}_{\perp}/\hbar \omega \cos(\Theta)
\right]$ in Eq.(\ref{A}), where $J_0(x)$ is the zero-order Bessel
function, $\tilde{t}_{\perp}=t_\perp J_0(k_Fd\tan(\Theta))$,
$t_{\perp}$ is the  out-of-plane hopping integral, and $\hbar k_F$
is the Fermi momentum.

\begin{figure}
\begin{center}
\includegraphics[angle=-90,width=0.50\textwidth]{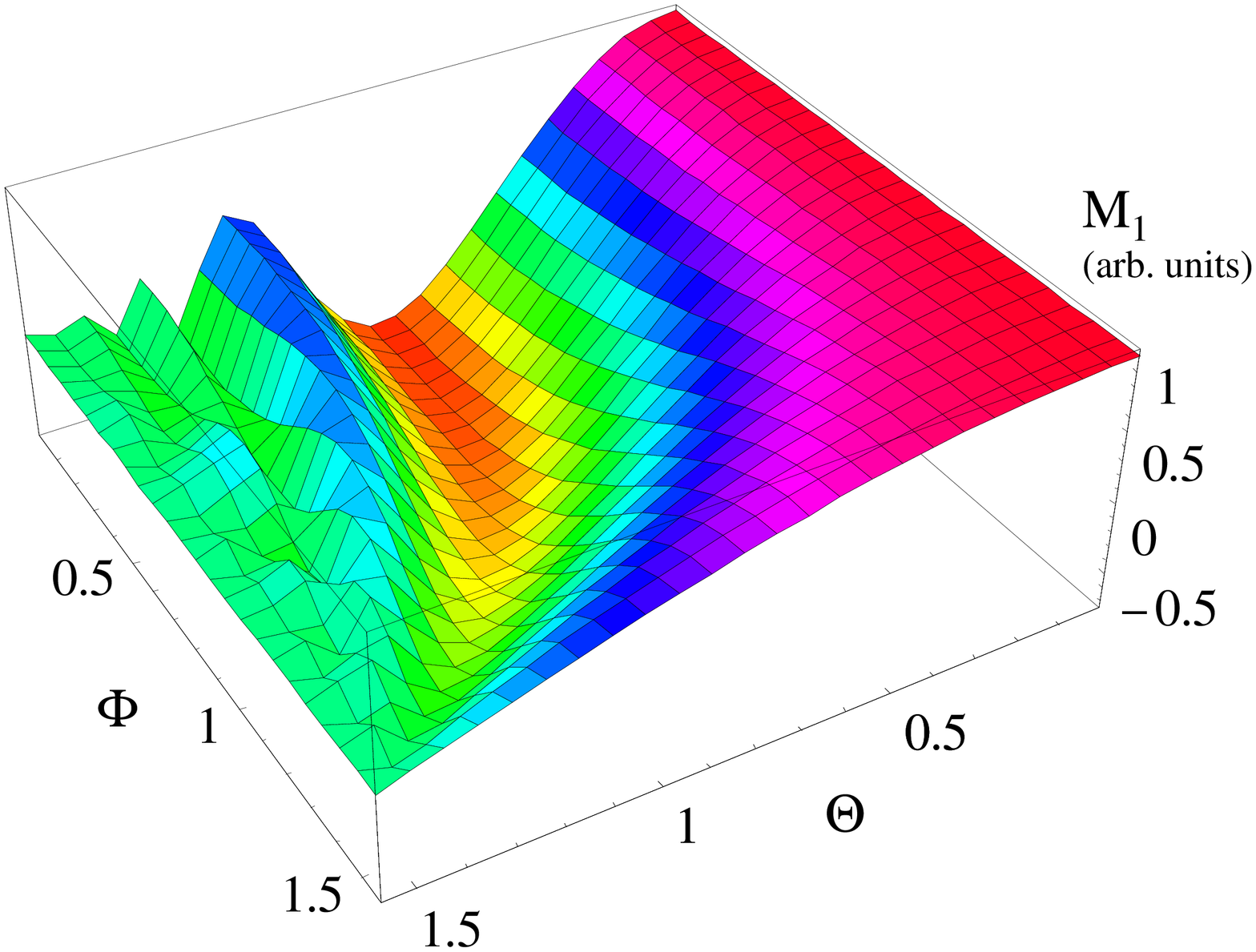}\\[0.2cm]
\includegraphics[angle=-90,width=0.50\textwidth]{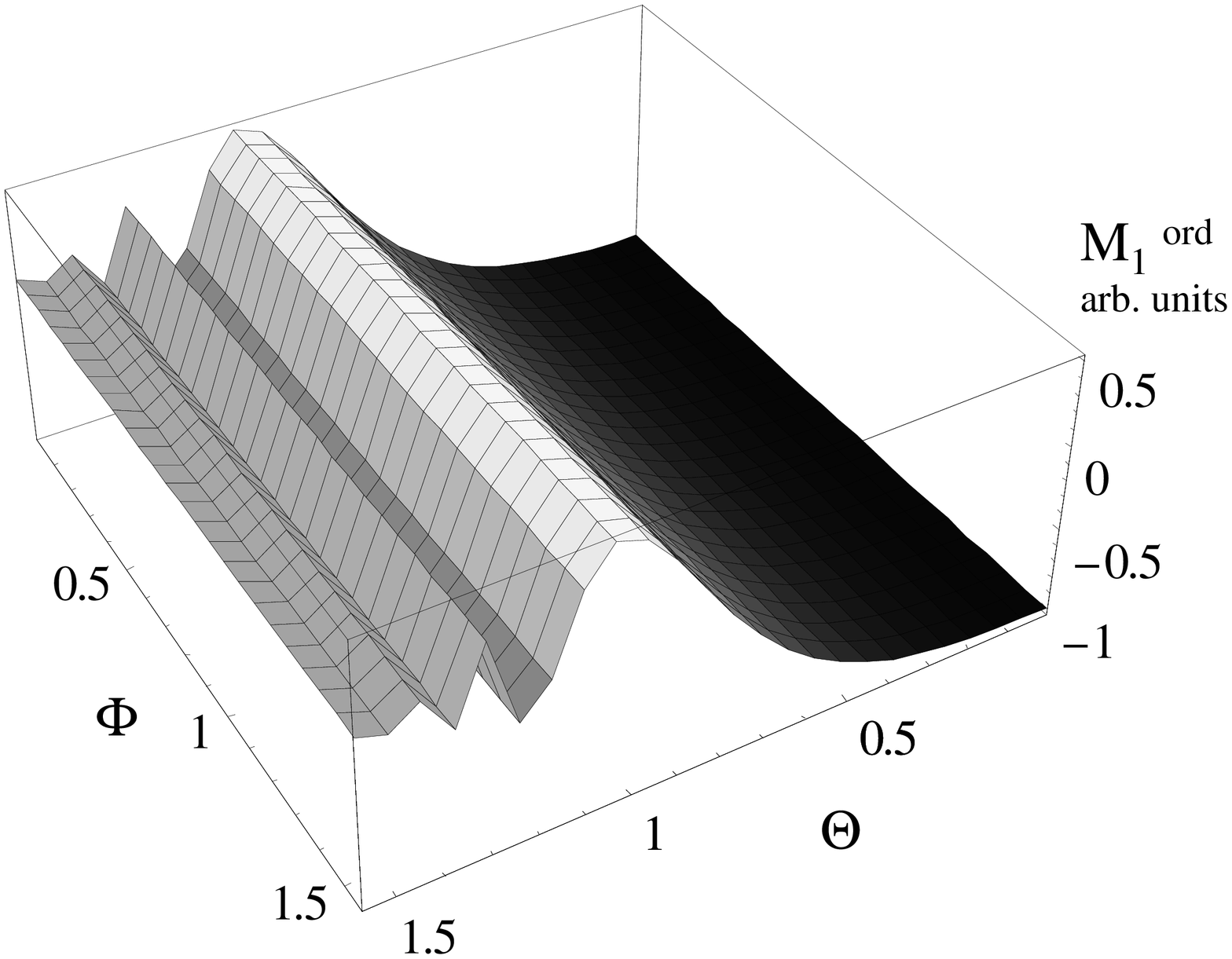}
\caption{dHvA first-harmonic amplitude as a function of
 the azimuthal in-plane angle from the magnetization direction,
 $\Phi$
and the polar angle from the out-of-plane direction, $\Theta$, in a
layered antiferromagnet (upper panel) compared with the first
harmonic amplitude in a nonmagnetic layered metal (lower panel) at
$T=\Gamma=0$, and $(m_xm_y)^{1/2}=m_e$.}
\end{center}
\end{figure}

As follows from  Eq.(\ref{M}) the essential anisotropy of the
$g$-factor causes a strong dependence of the oscillation amplitude
on the azimuthal in-plane angle of the field from the magnetization
direction, Fig.2a, which is absent in ordinary non-magnetic layered
metals, Fig.2b, where the magnetization amplitudes are found as
\begin{equation}
M^{ord}_r=A_r(\Theta)\cos \left[{\pi r (m_xm_y)^{1/2}\over {m_e \cos
(\Theta)}}\right] \label{M2}.
\end{equation}
The novel dependence on $\Phi$ and $\Theta$, Eq.(\ref{M}), is
extremely  pronounced at low temperatures (compare Fig.2 (upper
panel) and Fig.2 (lower panel)), as also shown in Fig.3 for some
fixed azimuthal angles.

\begin{figure}
\begin{center}
\includegraphics[angle=-0,width=0.45\textwidth]{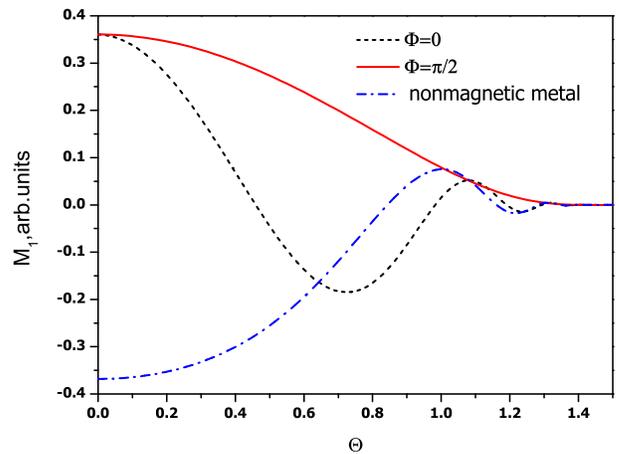}\\[0.2cm]
\caption{dHvA first-harmonic amplitude as a function of the polar
angle, $\Theta$, for two different azimuthal in-plane angles,
 $\Phi$,
 in a layered antiferromagnet  compared with the $\Phi$-independent
first harmonic amplitude in a nonmagnetic layered metal  at $T=0$,
$2\pi \Gamma=\hbar \omega$, $\gamma k_F=0.1$, and
$(m_xm_y)^{1/2}=m_e$.}
\end{center}
\end{figure}

 One can readily generalize our results to any shape of the Fermi surface, and
 calculate  corrections  to  amplitudes and  fundamental
 frequencies of higher order in $E_F/\Delta$ and in the
 magnetic field by applying the Lifshits-Kosevich quasi-classical
approximation \cite{kos}.  Within the approximation  dHvA
frequencies $F_{\pm}$ are determined by the extremal cross-section
areas, $S_{\pm}^{ext}$ of  two spin-split electron (or hole) Fermi
surfaces, $F_{\pm}=\hbar S_{\pm}^{ext}/2\pi e$. Following
Ref.\cite{mineev} one can expand the extremal cross-section area in
powers of the magnetic field, so that $F_{\pm}=F \pm \alpha B  +
\beta B^{2}\pm \epsilon B^{3}$. Here the second term describes the
Zeeman splitting of the bands with the anysotropic $g$-factor. It
does not  shift the frequency but affects the amplitude.
\begin{figure}
\begin{center}
\includegraphics[angle=-90,width=0.45\textwidth]{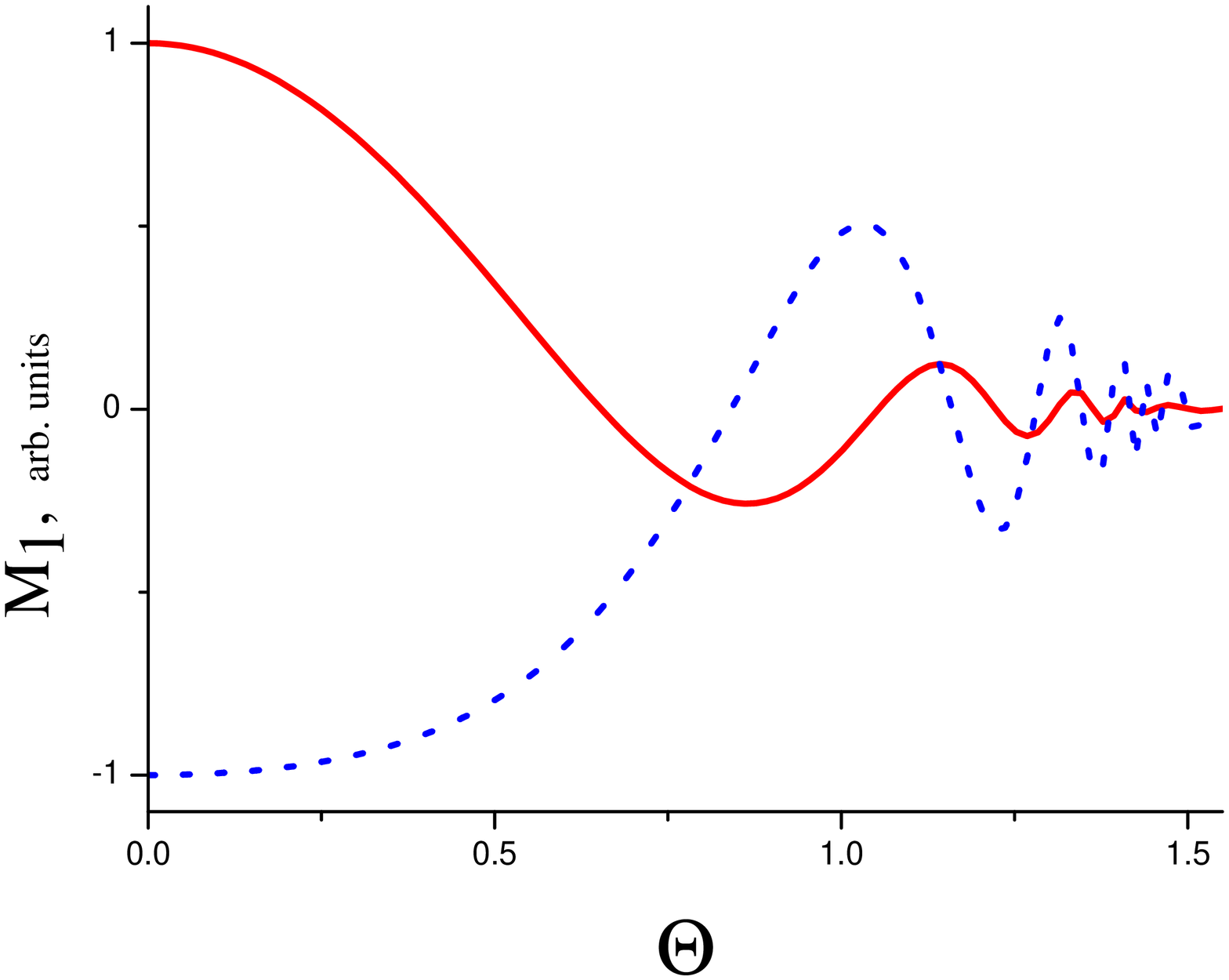}\\[0.2cm]
\caption{dHvA first-harmonic amplitude as a function of the polar
angle, $\Theta$,
 in a  disordered antiferromagnet (solid line)  compared with the
first harmonic amplitude in a nonmagnetic layered metal (dotted
line) at $T=\Gamma=0$, and $(m_xm_y)^{1/2}=m_e$.}
\end{center}
\end{figure}
The third term describes a small shift of the fundamental frequency,
$F$, depending on the magnetic field.  The last term describes a
small field-dependent correction to the $g$-factor. For example,
when the field is perpendicular to the magnetization,
$B_{\parallel}=0$, and the effective mass approximation  is applied
near $X$-point, $(\pi/2a, \pi/2a)$,  of the antiferromagnetic
Brillouin zone, one finds $F= \hbar k_{F}^{2}/2e$, $\alpha=
(m_xm_y)^{1/2} \gamma k_{F}/\pi m_e$, $\beta=
m_x^{3/2}m_y^{1/2}\gamma^{2}e/8m^2_e \hbar$, and $\epsilon=
m_x^{9/4}m_y^{3/4}\gamma^{3}e^2/12\pi m_e^{3} \hbar^2 k_{F}$ with $
\hbar k_F= [2(m_x m_y)^{1/2}E_F]^{1/2}$.   For an arbitrary field
direction  one obtains,  using Eq.(\ref{appr}) with $\gamma({\bf
k})=\gamma k_x \ll 1$,
 \begin{eqnarray}
&&{M_r\over{A_r(\Theta)}}= \cr \nonumber && \cos \left[{ 2r
\left[m_xm_y(cos^{2}(\phi)+\gamma^{2}k_F^2
 \sin^2(\phi))\right]^{1/2}E[\kappa(\phi)]\over{
m_e\cos(\Theta)}}\right].\\ \label{a}
\end{eqnarray}
Here $\phi$ is the angle between the magnetic field and the
magnetization, Fig.1,  $E(\kappa)$ is the elliptic integral of the
second kind, and
\[\kappa(\phi)=\Bigl [{\gamma^{2} k_{F}^{2} \sin^2(\phi)\over
{\cos^2{(\phi)}+\gamma^{2} k_{F}^{2}\sin^2{(\phi)}}} \Bigr ]^{1/2}.
\]
Taking $\gamma=0$ in Eq.(\ref{a}) one obtains Eq.(\ref{M}) since
$\cos^2(\phi)=\sin^2(\Theta) \cos^2(\Phi)$ and $E[0]=\pi/2$,
$E[1]=1$. The finite transverse spin-susceptibility, $\propto \gamma
k_F= (2E_F/\Delta)^{1/2}$, only slightly blurs the  strong
$\Phi$-dependence of the amplitudes, Fig.3,  if $E_F/\Delta \ll 1$.
For example, when the field is rotated in the plane perpendicular to
the magnetization axis $\mathbf{n}$ we have
\begin{equation}
M_r=A_r(\Theta)\cos \left[{2r(2m_xm_yE_{F}/\Delta)^{1/2}\over
{m_e\cos(\Theta)}}\right]
\end{equation}
 with a small transverse
$g$-factor. On the other hand if the magnetic field is rotating in
the $(z,\mathbf{n})$ plane, the angular dependence is quite
different,
\begin{equation}
M_r=A_r(\Theta)\cos\left[{\pi r
(m_xm_y)^{1/2}\tan(\Theta)\over{m_e}}\right],
\end{equation}
 as in Eq.(\ref{M}) with $\Phi=0$.

Real antiferromagnetic solids, like cuprates, could be disordered or
twinned, so that the magnetization direction ${\bf n}$ within the
plane is random. Nevertheless the dependence of dHvA amplitudes on
the polar angle, $\Theta$, remains rather unconventional. Indeed
averaging Eq.(\ref{M}) over all directions of $\Phi$ from zero to
$\pi/2$ yields
\begin{equation}
\langle M_r\rangle=A_r(\Theta)J_0 \left[{\pi r (m_xm_y)^{1/2}
\tan(\Theta) \over {m_e }}\right] \label{M3},
\end{equation}
which is distinguishably different from the amplitudes in a
nonmagnetic metal, Eq.(\ref{M2}), Fig.4.  There are
 known relations between oscillations in transport and thermodynamic
 quantities \cite{shoen}, at least in nonmagnetic substances. Relying
  on them, we expect the similar nontrivial angle dependences also in the SdH
magnetooscillations.

In summary, we have  derived the energy spectrum of electrons
(holes) doped into a two-dimensional antiferromagnetic insulator in
terms of all-neighbours hopping integrals of nonmagnetic lattice,
Eq.(\ref{general}), and  quantized it in the external magnetic field
of arbitrary direction. The peculiar dependence of dHvA/SdH
magneto-oscillation amplitudes on
 the azimuthal in-plane angle from the magnetization direction
and on the polar angle from the out-of-plane direction is found,
which could be instrumental as a sensitive probe of the
antiferromagnetic order in doped Mott-Hubbard, spin-density wave
(SDW), and conventional band-structure  insulators.

We greatly appreciate valuable discussions with Revaz Ramazashvili
and Iorwerth Thomas, and support of this work by EPSRC (UK) (grant
No. EP/D035589).

\end{document}